\documentclass[preprint,amsmath,amssymb]{revtex4}
\usepackage{graphicx}
\usepackage{dcolumn}
\usepackage{bm}
\begin{document}
\title{Vacuum energy and the spacetime index of refraction: A new synthesis}
\author{ M.~Nouri-Zonoz $^{a,b}$ \footnote{
Electronic address:~nouri@theory.ipm.ac.ir}}
\address{$^{a}$ Department of Physics, University of Tehran, North Karegar Ave., Tehran 14395-547, Iran. \\
$^{b}$ School of Astronomy and Astrophysics, Institute for Research in
Fundamental Sciences (IPM), P. O. Box 19395-5531 Tehran, Iran.}

\begin{abstract}
In $1+3$ (threading) formulation of general relativity spacetime  behaves analogous to a  medium with a specific  index of refraction with respect to the light propagation. Accepting the reality of zero-point energy, through the equivalence principle, we elevate this analogy to the case of virtual photon propagation in a quantum vacuum in a curved background spacetime. Employing this new idea one could examine the response of vacuum energy to the presence of a weak stationary gravitational field in its different quantum field theoretic manifestations such as Casimir effect and Lamb shift. The results are given explicitly for a Casimir apparatus in the weak field limit of a Kerr hole.
\end{abstract}
\maketitle
%%%%%%%%%%%%%%%%%%%%%%%%%%%%%%%%%%%%%%%%%%%%%%%%%%%%%%
\section{Introduction}
There was a time that the concepts of vacuum, space and time were mostly thought to belong to the philosophical realm and it is amazing that they all mathematically formulated in the language  of physics in
the beginning of the last century through the introduction of quantum mechanics and relativity.
Furthermore it
seems that the most challenging endeavor of physicists, a quantum theory of gravity, could only be achieved through
an appropriate conceptual unification of these fundamental concepts in a consistent mathematical setting.
After quantum mechanics there is no vacuum in the popular sense of an absolute emptiness or void and we believe in
an ever-fluctuating {\it quantum vacuum} and after general relativity the concepts of space and time merged to form a new concept leading to an intimate relation between geometry and matter encoded in the Einstein field equations.
Recently the so-called cosmological constant problem, which arises at the intersection of cosmology and
quantum field theory, has forced us to go back to basics and look at the roots of our standard models in more
detail and question all the original concepts, definitions and assumptions employed in the formulation of this problem. One of the main
concepts in the black list of those that look suspicious is the vacuum energy and its possible gravitational and cosmological implications. In this regard, the first question needing to be answered is whether or not-- and in what sense-- the vacuum energy gravitates?. The answer to this question will naturally shed light on the controversial issue of the reality of zero-point energy \cite{Jaffe}. Along the road to answering this question, it has been recently discussed how vacuum energy responds to the presence of a gravitational field by examining the force experienced by a Casimir apparatus in a {\it weak} gravitational field \cite{Fulling, Calloni1, Calloni2, Bimonte}. In \cite{Fulling}, considering the variation in the gravitational energy,
\begin{eqnarray}\label{energy}
E_g = -\frac{1}{2}\int d^3 x  h_{ab} T^{ab}
\end{eqnarray}
in a static, weak gravitational field ($g_{ab} = \eta_{ab} + h_{ab}$) represented by the Fermi metric 
\begin{eqnarray}\label{metric1}
g_{00} = -(1 +  2gz)\;\;\;\; , \;\;\;\; g_{\mu\nu} = \delta_{\mu\nu}\;\;\;\;, \;\;\;\; \mu,\nu = 1,2,3
\end{eqnarray}
(with $-z$ being the direction of gravity) the force on the Casimir energy per unit area of an apparatus oriented with a given angle with respect to the direction of the gravitational field is found to be
\begin{eqnarray}\label{force0}
{F} = -g {\cal E}_{Casimir} = g\frac{\pi^2 {\hslash}}{720 a^3 c}
\end{eqnarray}
with $g=\frac{GM}{R^2}$ the gravitatiobal acceleration, $a$ the plate separation and 
${\cal E}_{Casimir}=-\frac{c{\hslash}\pi^2}{720a^3}$ 
the famous (flat space) Casimir energy per unit area. Note that $T^{ab}$, due to the weak field limit, is set to be the flat, one-loop expectation value of electromagnetic energy-momentum tensor given by
\begin{eqnarray}\label{emt}
<T^{ab}> = -\frac{c\pi^2 {\hslash}}{180a^4}(\frac{1}{4}\eta^{ab} + {\hat x^a}{\hat x^b}).
\end{eqnarray}
Obviously the force (\ref{force0}) is independent of the orientation of the plates with respect to the gravitational field (as expected from the scalar nature of mass/energy) and because of its sign gives rise to a push upward on the apparatus. In simple words they have shown that the gravitational mass of the  {\it renormalized} Casimir energy is nothing but its energy itself, a result also reported (in two dimensions) earlier in \cite{Jaekel}.
It is also pointed out that if instead of (\ref{metric1}) one employs the weak field limit of the Schwarzschild metric, the same approach based on (\ref{energy}) gives different results in different coordinate systems, an obvious drawback which was  discussed to be rooted in the relation between coordinate increments and physical distances in different coordinate systems. We will get back to this important point later.\\ Here we are also interested in the interaction of vacuum energy with a weak gravitational field but through a quite different and novel approach which has far reaching consequences. To be more explicit, the scenario we are going to follow is this: in QED zero-point energy is related to the virtual photon propagation, such as in the famous cases of the Casimir effect and the Lamb shift, so we ask the question;
what are the consequences of treating such virtual photons capable of being affected by a gravitational field in the same way that the real photons are seemingly affected by the spacetime in which they propagate?. The way that virtual photons are affected by a weak gravitational field is described in our main assumption below:\\

{\bf Conjecture}:\\
{\it Accepting that the virtual photons do gravitate, through equivalence principle, their propagation should be influenced by the presence of a gravitational field in the same way that real photons are being influenced by a gravitational field. In the case of real photons, through the interpretation of modified Fermat's principle in a curved spacetime, this influence is encoded in their frequency change induced by the spacetime 
index of refraction.}\\ 

In principle this conjecture applies whenever we sum over virtual photon frequencies to calculate a quantum
vacuum effect in a weak gravitational field.
In this article, using the above  conjecture, we will consider the response of vacuum energy to the presence of a weak  {\it stationary} gravitational field. To study a specific example we consider the case of Casimir energy in the gravity field of a rotating object (Kerr spacetime) in its weak field limit. We will also briefly comment on the interaction of vacuum energy with a gravitatioal field in the context of variation in the Lamb shift of an atom in the presence of a gravitational field.
These studies, leading to consistent results expected on physical grounds, may have logically been taken as further evidence for the reality of zero-point energy.\\
%%%%%%%%%%%%%%%%%%%%%%%%%%%%%%%%%%%%%%%%%%%%%%%%%%%%%%%%%%%%%%%%%%%%
\section{Spacetime as a refractive medium}
Starting with the geometrical side of our assumption which is the assignment of an index of refraction to a spacetime, we introduce the $1+3$ (threading) formulation of spacetime decomposition which leads to the so-called Gravitoelectromagnetism (GEM) and the quasi-Maxwell form of the Einstein field equations \cite{Landau, Lynden}. In this approach the metric of a stationary spacetime is written in the following form,
\begin{eqnarray}\label{metric}
ds^2 = g_{00}(dx^0 -{A_\alpha}dx^\alpha)^2 -dl^2
\end{eqnarray}
where $A_{\alpha}=-\frac{g_{0\alpha}}{g_{00}}$ is the so-called gravitomagnetic potential and 
\begin{eqnarray}
dl^2 = \gamma_{\alpha\beta}dx^{\alpha}dx^{\beta} = (-g_{\alpha\beta} + \frac{g_{0\alpha}g_{0\beta}}{g_{00}})dx^\alpha dx^\beta\;\;\;\alpha, \beta = 1,2,3
\end{eqnarray}
is the spatial length element in terms of the three-dimensional spatial metric $\gamma_{\alpha\beta}$. In this context one can attribute, through Fermat's principle in a curved background, the following index of refraction to a stationary spacetime \cite{Landau,EFE}
\begin{eqnarray}\label{RI}
n = \frac{1}{\sqrt{g_{00}}} + A_{\alpha}\frac{dx^{\alpha}}{d l}.
\end{eqnarray} 
It can also be written in the following form \footnote{This so-called {\it spacetime index of refraction} being  a purely geometrical entity is different from the {\it vacuum refractive indices} assigned with nontrivial vacuua such as those induced by vacuum polarization effects in quantum gravitational optics \cite{Drummond}, in Casimir vacuum \cite{Scharn} or quantum gravitational effects based on spacetime foam model \cite {Amelino}. In \cite{Ahmadi} these two concepts are combined by calculating the contribution of the vacuum polarization effects on the  spacetime refractive index in the context of semi-classical gravity and it is shown that the local index of refraction is modified by a term which is proportional to the null cone modification itself.}
\begin{eqnarray}
n(x^\mu,{\dot{x}}^\mu) = \frac{1}{\sqrt{g_{00}}} + A_{\alpha}\frac{dx^{\alpha}}{d \tau}\frac{d \tau}{dl}=\frac{1}{\sqrt{g_{00}}} + \frac{1}{c}A_{\alpha}v^{\alpha},
\end{eqnarray} 
in which  $v^\alpha = \frac{dx^{\alpha}}{d \tau}$ is the 3-velocity of a point moving in the field. It should be noted that $\frac{dl}{d\tau}$  is set equal to $c$, the velocity of light, according to the definition of the spatial distance element. This is a delicate fact and of utmost importance in what follows since the spatial distance is measured using the proper time of an observer who sends and receives light signals (on the same path) to a nearby observer \cite{Landau} and consequently leads us to coordinate independent results in our approach.
Physically interpreted, attributing an index of refraction to a spacetime means that if  the allowed photon wavelengths are somehow fixed, e.g., through the presence of boundaries, then the effect of refractive index is that it changes the photon frequencies from $\omega_k$ to $\frac{\omega_k}{n}$. In the case of a static gravitational field such as the Schwarzschild metric the index of refraction reduces to the simple form $n_{static} = \frac{1}{\sqrt{g_{00}}}$. We note that in this case the index of refraction is given by the same quantity accounting for the redshift in a constant spacetime (in the sense that the world time frequency $\omega_0$ and the proper time frequency $\omega$ are related by $\omega = \frac{\omega_0}{\sqrt{g_{00}}}$ \cite{Landau}). Of course, these are distinctly different concepts and there is no such coincidence in the stationary case. The redshift accounts for the frequency shift between the frequencies of light emmitted and those received in two different points in {\it a gravitational field} but the index of refraction, in the present context, tells us how the propagation of light differs in the presence of a gravitational field {\it compared } to the case in its absence (i.e in the flat case) \footnote{That is why the unjustified calculations based on redshifted modes inbetween the plates in \cite{Calloni1} has led to the wrong answer in the static case and again it is not clear why all modes propagating both normal and parallel to the plates, unlike in \cite{Calloni1}, are redshifted by the same factor in \cite{Calloni2}. In any case it is clear that using the same approach based on redshifted modes will not work for stationary spacetimes.}.
%%%%%%%%%%%%%%%%%%%%%%%%%%%%%%%%%%%%%%%%%%%%%%%%%%%%%%%%%%%%%%%%%%%%%%%%%%
\section{Vacuum energy in a gravitational field}
Using the above ideas we calculate the gravitational force on {\it finite} Casimir energy by subjecting it to a stationary gravitational field represented by a metric of general  form (\ref{metric}) and the corresponding index of refraction (\ref{RI}).
As pointed out in the introduction, since we work in the weak field limit there is no need for the full weaponry of qunatum field theory in curved spacetime and the effect of the gravitational field can be found mode-by-mode using the flat space result for Casimir energy employing the spacetime refractive index. To this end, we first note that in the absence of the gravitational field the {\it unrenormalized} zero-point energy per unit area between plates is given by \cite{Bordag},
\begin{eqnarray}
{\cal E}_{0}= -\frac{c{\hbar}}{2}\int_0^\infty dk_1 d k_2\sum_{m=-\infty}^ {\infty}\sqrt{{k_1}^2 + {k_2}^2 
+ (\frac{m\pi}{a})^2}
\end{eqnarray}
However in the presence of the gravitational field, every mode's frequency $\omega_k$ is changed to $\frac{\omega_k}{n}$ and so the unrenormalized vacuum energy is given by,
\begin{eqnarray}
{\cal E}_{0}^{grav}= -\frac{c{\hbar}}{2}\int_0^\infty dk_1 d k_2\sum_{m=-\infty}^ {\infty}\sqrt{\frac{({k_1}^2 + {k_2}^2 +(\frac{m\pi}{a})^2)}{n(x^\mu , {\dot{x}}^\mu)^2}}= \frac{{\cal E}_0}{n(x^\mu , {\dot{x}}^\mu)}
\end{eqnarray}
Being dependent on the position of the apparatus and its velocity, the index of refraction is just a multiplication factor and so
neglecting the divergent contribution to the vacuum energy, the {\it renormalized} zero-point energy in the presence of the gravitational field is given by  ,
\begin{eqnarray}
{\cal E}_{0,ren}^{grav}= \frac{{\cal E}_{0,ren}}{n(x^\mu , {\dot{x}}^\mu)} = \frac{{\cal E}_{Casimir}}{n(x^\mu , {\dot{x}}^\mu)}
\end{eqnarray}
The above results for the frequency and energy changes in a Casimir apparatus in a weak gravitational field are  direct consequences of applying our main assumption and one may not find them satisfactory. As a partial proof  of our  conjecture, using QFT in a static curved background, in an accompanying paper \cite{Nouri3} we have calculated the Casimir effect for a  massless scalar field in a weak gravitational field and it is shown that it leads to the above proposed changes in the virtual particle frequencies and  Casimir energy in terms of the space time index of refraction. Using the above expression for the Casimir enegy the gravitational force on the apparatus is given by
\begin{eqnarray}
F_\mu=(-\frac{\partial}{\partial x^\mu}+\frac{d}{dt}\frac{\partial}{\partial {\dot{x}}^\mu}){\cal E}_{0,ren}^{grav}= {\cal E}_{Casimir} (-\frac{\partial}{\partial x^\mu}+\frac{d}{dt}\frac{\partial}{\partial {\dot{x}}^\mu})(\frac{1}{n(x^\mu , {\dot{x}}^\mu)})
\end{eqnarray}
Now it is an easy task to show that in the weak field limit, where $n(x^\mu ; {\dot{x}}^\mu)$ reduces to
\begin{eqnarray}
n(x^\mu,{\dot{x}}^\mu) \approx 1 - \Phi + \frac{v^\mu A_\mu}{c}
\end{eqnarray} 
(with $g_{00} = e^{2\Phi}$) the force is given by \footnote{Note that, with respect to the gravitational field, the apparatus is taken as a point particle. In other words, due to the weak field limit, the variation of the spacetime metric inside the cavity is negligible so that the index of refraction and the force are calculated at the center of the cavity.} 
\begin{eqnarray}\label{force}
{\bf F} = {\cal E}_{0,ren}({\bf E}_g + \frac{\sqrt{g_{00}}}{c}{\bf v}\times {\bf B}_g)
\end{eqnarray}
where ${\bf E}_g = -\nabla ln \sqrt{g_{00}} = -\nabla \Phi$ and ${\bf B}_g =\nabla \times {\bf A}$ are the gravitoelectric and gravitomagnetic fields associated with the underlying spacetime \cite{Lynden,Nouri}.
Note that this is quite a nontrivial result  which is based on the assumption of the interaction of virtual photons with the underlying weak gravitational field through its refractive index. Comparing this with the gravitoelectromagnetic Lorentz force on a particle of mass $m$ in the weak field limit of a stationary spacetime \cite{Landau,Lynden,Rindler},
it is seen that the finite Casimir energy acts like any other energy/mass in response to the gravitational field. Indeed  going through the details in the weak field limit for a slowly rotating object with angular momentum $J=Ma$ presented by the weak field and slow rotation limit of the Kerr metric (in Boyer-Lindquist coordinates),
\begin{eqnarray}
ds^2 = (1-\frac{2M}{r}) dt^2 - (1+\frac{2M}{r})(dr^2 +r^2 d\Omega^2) + \frac{4J}{r} \sin^2 \theta dt d\phi
\end{eqnarray}
the above force for ${\bf v}=v_\phi{\bf \phi_0}= r \omega sin\theta {\bf \phi_0}$ at constant $r$ and $\theta$, reduces to
\begin{eqnarray}
{\bf F}\approx {\cal E}_{0,ren}\{(-\frac{M}{r^2} + r\omega^2 \sin^2\theta){\bf {{r}_0}} - r\omega^2\sin(2\theta) {\bf {\theta}_0}\}
\end{eqnarray}
in which $\omega=\frac{d\phi}{dt}\approx\frac{2J}{r^3}$ is the  angular velocity of the  ZAMOs (as well as the Casimir apparatus due to dragging effect) at radius $r \gg a, m$  measured by the observers at infinity.
Now the force depends not only on the mass of the object but also on how much of the object's rotation (which by principle of equivalence is also a source of inertia) is seen or felt by the Casimir energy. Two cases are of interest, first the force on the axis of rotation ${\bf F}_{axis}\approx -{\cal E}_{0,ren}\frac{M}{r^2}{\bf {{r}_0}}$ which as expected has no contribution from rotation and secondly in the equatorial plane $\theta =\frac{\pi}{2}$ where it reduces to
\begin{eqnarray}
{\bf F}_{equat.}\approx {\cal E}_{0,ren}(-\frac{M}{r^2} + r\omega^2 ){\bf r_0}
\end{eqnarray}
which, by the sign of ${\cal E}_{0,ren}$, apart from a push in the {\it outward} radial direction contributes a {\it centripetal} force originated from the hole's rotation. Therefore using our physically motivated synthesis it is shown that the Casimir energy couples to both the mass and rotation induced inertia like any other energy/mass.
%%%%%%%%%%%%%%%%%%%%%%%%%%%%%%%%%%%%%%%%%%%%%%%%%%%%%%%%%%%%%%%%%%% 
\section{Discussion and summary}
As pointed out in the introduction the approach based on the variation in the gravitational energy \cite{Fulling} 
gives different results for the force exerted on the Casimir energy in different coordinate systems. So one needs to specify which coordinate system should be taken as the prefered or physical one and why?. Looking at the approach employed above one could see the reason why in this case we have not faced the same problem and that is the choice made for the definition of the spatial/physical distance $dl$  in $1+3$ formalism which is based on the proper time measured by an observer sending and receiving light signal to a nearby observer. In other words we have employed  the spacetime metric in its general $1+3$ form (\ref{metric}) which led to the coordinate independent definition of spacetime index of refraction and its weak field limit.
In summary regarding the approach based on the spacetime index of refraction leading to the force on vacuum energy (\ref{force}), we note the following points: \\
1-There is no ambiguity as to which coordinate is used \cite{Fulling} and there is no dependence on the orientation of the apparatus as expected from the equivalence principle and the scalar nature of mass.\\
2-It is applicable to any stationary spacetime and in principle it can be employed to find out coupling to gravity of other vacuum energy manifestations which are calculated through summation over virtual photon frequencies.\\
Related to this last point, it is quite intersting to find out that Feynman has actually used a similar argument based on the  index of refraction of a dilute gas of $N$ atoms in a box of volume $V$ to interpret the Lamb shift as a change in the zero-point energy due to the index of refraction introduced by the presence of the atoms without a need for mass renormalization \cite{Feynman}. Indeed, using Feynman's interpretation, as a further justification/test of the above synthesis one should be able to examine the coupling to gravity of vacuum energy in the case of an atom in a weak gravitational field through the change in its Lamb shift. A detailed study of this issue will appear elsewhere \cite{Nouri2}.
%%%%%%%%%%%%%%%%%%%%%%%%%%%%%%%%% 
\section *{Acknowledgments}
The author would like to thank Gary Gibbons for useful discussions. He also thanks Center of Excellence on the Structure of Matter and particle interactions  of the Unversity of Tehran for partial support. This project is also supported by a grant from  the research council of the University of Tehran.
\pagebreak
%%%%%%%%%%%%%%%%%%%%%%%%%%%%%%%%%%%%%%%%%


\begin{thebibliography}{30}
\bibitem{Jaffe} R. L. Jaffe, Phys. Rev. D72 (2005) 021301.
\bibitem{Fulling} S. A. Fulling et al., Phys. Rev. D {\bf 76} 025004 (2007).
\bibitem{Calloni1} E. Calloni et al., Int. J. Mod. Phys. A, 17 (2002) 804.
\bibitem{Calloni2} E. Calloni et al., Phys. Lett. A,  {\bf 297} 328 (2002).
\bibitem{Bimonte} G. Bimonte et al., Phys. Rev. D, {\bf 74} (2006) 085011; Erratum ibid. D {\bf 75}, 049904(2007); Erratum ibid. D {\bf 75}, 089901 (2007); Erratum ibid. D {\bf 77}, 109903 (2008).
\bibitem{Jaekel} Jaekel, M. T. and Reynaud, S., J. Phys. I France 3 (1993) 1093-1104.
\bibitem{Landau} Landau, L. D. and  Lifshitz, E. M., {\it The classical theory of fields}, Pergamon Press, 1975, p 254.
\bibitem{EFE} Schneider, P., Ehlers, J. and Falco, E. E.,{\it Gravitational lenses}, Springer 1999, P 104.
\bibitem{Lynden} Lynden-Bell, D. and Nouri-Zonoz, M., Rev. Mod. Phys. 70, 427 (1998).
\bibitem{Drummond} Drummond, I. T. and Hathrell, S. J., Phys. Rev. D {\bf 22}, 343 (1980); Shore, G. M., Nucl. Phys. B {\bf 778}, 219 (2007).
\bibitem{Scharn} Scharnhorst, K., Phys. Lett. B {\bf 236}, 354 (1990); Barton, G. and Scharnhorst, K., J. Phys. A {\bf 26}, 2037 (1993).
\bibitem{Amelino} Amelino-Camelia, G., et al., Int. J. Mod. Phys. A, {\bf 12}, 607 (1997); J. Ellis et al. arXiv:0804.3566 [hep-th].
\bibitem{Ahmadi} Ahmadi, N. and Nouri-Zonoz, M., 05 JCAP 2008.
\bibitem{Nouri3} Nazari, B. and Nouri-Zonoz, M., ArXiv:1003.0614.
\bibitem{Nouri} Nouri-Zonoz, M., Gravomagnetic monopoles, Ph.D thesis, University of Cambridge, 1998.
\bibitem{Bordag} M. Bordag, U. Mohideen, V.M. Mostepanenko; Phys. Rept. 353:1-205, 2001.
\bibitem{Rindler} Rindler W., {\it Relativity: Special, General and cosmological}, 2-nd ed., Chapter 9, Oxford university press, 2006.
\bibitem{Feynman} Feynman R. P.,{\it The present status of quantum electrodynamics}, in the {\it Quantum theory of fields}, ed. Stoops, R., Wiley interscience, 1961; Simplified version of Feynman's argument is given in: Power, E. A., Am. J. Phys. {\bf 34} 516 (1966) see also  Milonni, P. W., {\it The Quantum Vacuum}, Academic Press, 1994.
\bibitem{Nouri2} Nazari, B. and Nouri-Zonoz, M, in preparation.
\end{thebibliography}
\end{document}